# Imaging propagating terahertz collective modes in two-dimensional semiconductor double layers

Andrew T. Pierce[1,2,3]*[‡], Chirag Vaswani[4]*, Dimitri Pimenov[1], Sihong Xu[1,5], Kenji Watanabe[6], Takashi Taniguchi[7], Erich Mueller[1], Debanjan Chowdhury[1], Kin Fai Mak[1,2,3,4][‡] and Jie Shan[1,2,3,4][‡]

[1]*Laboratory of Atomic and Solid State Physics, Cornell University, Ithaca, NY, USA*
[2]*Kavli Institute at Cornell for Nanoscale Science, Ithaca, NY, USA*
[3]*Max Planck Institute for the Structure and Dynamics of Matter, Hamburg, Germany*
[4]*School of Applied and Engineering Physics, Cornell University, Ithaca, NY, USA*
[5]*Department of Physics, The Hong Kong University of Science and Technology, Hong Kong, China*
[6]*Research Center for Electronic and Optical Materials, National Institute for Material Science, 1-1 Namiki, Tsukuba 305-0044, Japan*
[7]*Research Center for Materials Nanoarchitectonics, National Institute for Material Science, 1-1 Namiki, Tsukuba 305-0044, Japan*

*These authors contributed equally to this work.
‡Corresponding authors' emails: atp66@cornell.edu, jie.shan@mpsd.mpg.de, kin-fai.mak@mpsd.mpg.de

**Two-dimensional transition metal dichalcogenide (TMD) semiconductors exhibit a wide range of novel phenomena at millielectronvolt (terahertz-frequency) energy scales, including superconducting and correlation-induced insulating gaps that are frequently accompanied by symmetry breaking. However, due to the subwavelength dimensions and the often low conductivities of these systems, their intrinsic THz plasmons and meV-scale excitation gaps are difficult to access experimentally. Here we report an optical readout method that can image propagating THz-frequency collective modes in real time. The method relies on a strong coupling between the optical polarons of monolayer TMD semiconductors and the local THz fields in a waveguide, which enables us to image THz plasmons with micron scale spatial resolution and determine their propagation group velocities. Moreover, at finite magnetic fields, we observe coherent cyclotron oscillations resulting from Landau level repopulation induced by the THz field. Our findings provide a new near-field platform for probing collective excitations in strongly correlated two-dimensional semiconductors and enable "all-photonic" TMD-based architectures for time-domain THz plasmonics and optoelectronics.**

**Main**

Van der Waals (vdW) heterostructures based on two-dimensional materials, including moiré superlattices based on graphene and transition metal dichalcogenide (TMD) semiconductors, frequently exhibit complex low-temperature phase diagrams[1] including superconductors, magnets, integer and fractional Chern and topological insulators and more. Many of these phases appear at temperatures of a few tens of kelvin and below, suggesting that spectroscopy in the THz frequency range may effectively probe their electronic properties and collective dynamics. However, the small sample size characteristic of vdW heterostructures necessitates on-chip[2–9] and near-field[10–16] techniques. These techniques generally require careful impedance matching, posing a problem for low-conductivity materials like heavy-electron moiré systems and TMD semiconductors, especially in the presence of disorder.

In this work we image THz-frequency collective excitations on the Fermi surface of TMD semiconductors by utilizing their strong coupling to optical polarons, i.e. optically excited excitons that can effectively polarize a Fermi liquid. The method allows us to visualize the dynamics of the collective excitations at micron length scales and at picosecond time scales. At zero magnetic field, we observe plasmons propagating in real space which, by virtue of the materials' large kinetic inductance, exhibit substantial confinement of order $\lambda/100$ ($\lambda$ is the wavelength in free space). In an applied magnetic field, we observe pronounced oscillations corresponding to quantum-coherent Landau level repopulation. Our method circumvents the impedance matching problem in direct on-chip THz transmission measurements on low-conductivity materials. Our results also open the door to new means of studying and controlling THz electrodynamics in two-dimensional semiconductors, and for studying polaron dynamics in a regime not accessible to ultracold gas systems that host only charge-neutral polarons.

**Excitonic sensing of local electrodynamics in WSe$_2$ enabled by on-chip THz generation**

Our experimental setup and vdW device geometry are shown schematically in Figs. 1a-b respectively (see Methods). A transmission line situated on an undoped silicon substrate hosts a vdW system of interest, here consisting of hBN-encapsulated and graphite-contacted WSe$_2$ and MoSe$_2$ monolayers (Fig. 1b). In this work, we hole dope the WSe$_2$ layer and electron dope the MoSe$_2$ layer; for a given doping, to maintain charge neutrality, the electron density on MoSe$_2$ is equal to the hole density on WSe$_2$. We probe the optical spectrum of WSe$_2$ near its 1$s$ exciton resonance, depicted in Fig. 1c, as the WSe$_2$ hole density is varied. The spectrum exhibits the charged-exciton phenomenology typical of monolayer TMDs[17,18], consisting of the so-called attractive and repulsive polaron branches[19], respectively, redshifted and blueshifted from the 1s exciton (labels "AP" and "RP", Fig. 1c). Sharp resonances attest to the high quality of the device.

To study the electronic response of WSe$_2$ to an applied THz field, we couple picosecond-timescale transient voltage pulses to the transmission line via on-chip photoconductive switches (see Fig. 1a and Methods), resulting in AC electric fields with strength estimated to be on the order of a few kV/cm oriented predominantly in the plane (see Methods). The optical response of WSe$_2$ to the applied THz field is characterized by the differential reflectance $\Delta R(E, t)/R$, defined as $(R_{on}(E, t)–R_{off}(E))/R_{off}(E)$, where $R_{on}(E)$ and $R_{off}(E, t)$ are the reflectance spectra with and without the THz field, respectively. Fig. 1d depicts $\Delta R/R$ as a function of hole density in WSe$_2$ at a fixed delay of 0.8 ps after the pulse. $\Delta R/R$ is found to be negligible at low densities and at charge neutrality (Fig. 1c), suggesting that, consistent with previous far-field THz-pump optical-probe studies[13,14], the THz field mainly addresses the charged rather than neutral exciton resonances. The observations confirm that the coupling between the optical resonances and the THz field depends crucially on the existence of a Fermi surface in the WSe$_2$ layer.

To examine the dynamical response to the THz field, we present in Figs. 1f-h measurements of $\Delta R/R$ at three different hole densities as a function of time delay between the THz and optical pulses near the THz injection location (see Fig. 1a). The magnitude of $\Delta R/R$ for the polaron branches follows the equilibrium density dependence of the optical spectra (Fig. 1c), exhibiting a transfer of optical spectral weight from the repulsive polaron branch to the attractive polaron branch with increasing hole density. Intriguingly, the arrival time of the peak value of $\Delta R/R$ (labeled with arrows in Fig. 1f-h), decreases monotonically as $n_{hole}$ increases from $4.1 \times 10^{11}$ cm$^{-2}$ to $1.4 \times 10^{12}$ cm$^{-2}$. Overall, these observations confirm the existence of a giant optoelectronic field effect, and the long density-dependent delays in the response compared to the pulse duration call for a closer examination of the high-frequency electrodynamics of the two-dimensional system.

**Visualizing coherent plasmonic wavepacket propagation at zero magnetic field**

The density dependence and tunable times-of-flight of the $\Delta R/R$ features suggest the response originates from electromagnetic propagation along the WSe$_2$ layer, and we therefore perform spatially resolved measurements of $\Delta R/R$ as a function of position $z$ between the electrodes of the transmission line (Fig. 2a). As the position of the probe beam is moved from $z=2$ μm to $z=12$ μm (Fig. 2b), and with $n_{hole}$ fixed at $4.1 \times 10^{11}$ cm$^{-2}$, the arrival time of the pulse is observed to increase monotonically, by about 3 ps, as the distance from the injection point increases by approximately 9 μm. We emphasize that none of these features appear in direct measurements of the THz pulse transmitted through the device along the transmission line (Extended Data Fig. 1), which instead show only a weak suppression of a few percent of the signal at low frequencies. This discrepancy arises from difficulty achieving impedance matching between the vdW heterostructure and the transmission line and highlights the sensitivity of our optical-readout scheme to local dynamics. We thus conclude that the measured $\Delta R/R$ signal reflects the propagation in real space of a mode excited by the THz field.

To compare the behavior of the propagating mode at different densities, we integrate the measured $|\Delta R(E, t)/R|$ spectra with respect to photon energy at each delay time and normalize by the peak value to account for position-dependent variation of the 1s contrast (see Methods). This procedure yields one-dimensional time series denoted $r(t)$ at each carrier density and position.

Examples are shown in Fig. 2c. Comparison of a measurement of $r(t)$ at a position close to the injection point with the input THz pulse reveals that, at high densities and on short time scales, $r(t)$ closely follows the input pulse (Fig. 2d), thus validating $r(t)$ as a metric for local THz response. In Figs. 2e-g, we present position-dependent measurements of $r(t)$ at $n_{hole}$=4.1×10$^{11}$ cm$^{-2}$, 8.9×10$^{11}$ cm$^{-2}$ and 1.4×10$^{12}$ cm$^{-2}$ respectively. Consistent with the observations in Fig. 1, we find that the velocity of the mode depends strongly on the carrier density: in the accessible range of densities, the mode propagates with apparent speeds of approximately 1-10 microns per picosecond (~$c$/100), substantially slower than the speed of light on the silicon substrate (~$c$/2.5, see Methods). As is well-known, geometric confinement alone cannot suppress the speed of light; instead, the dependence on density suggests that the AC conductivities of the TMD layers are responsible for the low propagation velocity. Thus, we identify the propagating modes as plasmon-polaritons (PPs) arising from the interaction between the light field and collective oscillations of the Fermi sea[20–27]. In this picture, the THz field changes the local electronic density, which via the gate dependence in Fig. 1c results in a transient modification of the optical spectra.

For a two-dimensional planar interface with complex sheet conductance $\sigma(\omega)$ embedded in a dielectric medium with effective permittivity $\varepsilon_{eff}$, plasmon propagation is described by a complex wavevector $q$ that, at sufficiently high frequencies and neglecting loss in a Drude description, is mainly determined by the kinetic inductance $L_k=m^*/ne^2$ according to $q=2\varepsilon_{eff}\omega^2 L_k$[21,28] (see Methods). For the case of WSe$_2$, the expected kinetic inductance exceeds the Faraday inductance, which is set by $\mu_0$, by around two orders of magnitude at the densities studied in Fig. 2 (see Methods). Such a giant enhancement of the mode inductance is superficially consistent with the sizable suppression of the speed of light inferred from Figs. 2e-g, highlighting the enormous degree of electromagnetic confinement exhibited by plasmons in TMD semiconductors.

Propagation in a real device is complicated by the presence of multiple dielectric layers, including boron nitride and silicon oxide, as well as the conductive MoSe$_2$ layer. Quantitative comparison with the experiment therefore requires us to calculate all allowed modes $\omega(q)$ (see Methods) in a realistic layered structure (Fig. 3a and inset). In the presence of two conductive interfaces, two bound surface modes are expected, corresponding to configurations (Fig. 3b) in which the fields and induced surface charges on the two layers are (panel i) in phase, referred to as the optical mode, and (panel ii) out of phase, referred to as the acoustic mode[29,30]. The modes' group velocities $v_{group}$ are shown in Fig. 3c as a function of $\omega$ over the same range of densities. For all densities and over the entire accessible frequency range, the group velocity of the optical mode greatly exceeds that of the acoustic mode by approximately a factor of 10. Crucially, in all cases, the measured velocities closely agree with the expected group velocities near the center of the pulse spectral weight for the optical mode, but never fall within the possible range of velocities for the acoustic mode. These characteristics unambiguously identify the propagating feature identified Fig. 2 as a broadband wavepacket of optical plasmons (panel iii, Fig. 3b).

While the mode spectra are sufficient to describe the dynamics in the steady state, they do not by themselves describe the impulse response needed to fully model the propagating transient. We therefore perform numerical calculations using the finite-difference time domain (FDTD) method in which we assume Drude-form conductivities for WSe$_2$ and MoSe$_2$, enabling us to extract the expected propagation velocity for an arbitrary input pulse (see Methods). In Fig. 3d, we compare the experimental propagation speeds with a model calculation including typical masses[31–34] and scattering times reported for MoSe$_2$ and WSe$_2$ (see Methods). We note that the simulated velocities are consistent with the measured values for $n_{hole}$>0.5×10$^{12}$ cm$^{-2}$, but over-estimate the propagation speed for lower densities by as much as a factor of 2 at $n_{hole}$=0.2×10$^{12}$ cm$^{-2}$ (where the uncertainty of the velocity extraction is the smallest, see Methods). We note that several theoretical approaches[35,36] report evidence for an effective mass enhancement, or even divergence, approaching the metal-insulator transition in a homogeneous electron gas. Alternatively, non-Ohmic and/or nonlinear contributions to the conductivity[37], if present, would alter the propagation velocity. Further experiments will be required to confirm the low-density velocity suppression effect and explore its underlying mechanisms. In sum, our findings demonstrate that the strong exciton-THz plasmon coupling reported here provides a powerful means of imaging electromagnetic propagation in vdW systems and raises the possibility of probing general THz spectral features associated with their electronic degrees of freedom.

**Coherent cyclotron resonance oscillations revealed by excitonic sensing**

The application of an out-of-plane magnetic field to a two-dimensional electron system radically alters its response to an applied THz field due to the quantization of its energy spectrum into Landau levels. Fig. 4a depicts $\Delta R(E, t)/R$ measured as a function of hole density near time zero ($t$=−0.2 ps). Analogous to prior studies of the equilibrium optical reflectance[38], we observe sharp modulations in $\Delta R/R$ near the repulsive polaron branch associated with the incompressible states formed at integer Landau level filling factors. These features are substantially more visible in the $\Delta R/R$ measurement as compared with the equilibrium reflectance (Extended Data Fig. 2), especially at low densities and for the attractive polaron branch[38,39], underscoring the sensitivity of our scheme to changes in the ground state of the system.

To investigate the dynamical interplay between the THz field and the Landau-quantized spectrum, we examine in Fig. 3b the delay dependence of $\Delta R(E, t)/R$ measured at $n_{hole} \approx 4.1×10^{11}$ cm$^{-2}$ and $B$=9 T. Strikingly, the $\Delta R/R$ spectra exhibit coherent oscillations in both the exciton and polaron branches with a period of approximately 1.4 ps, consistent with an optical transition at a photon energy of approximately 3 meV; notably, no such features appear in direct measurements of the THz pulse transmission (Extended Data Fig. 1d-f). The appearance of this behavior at finite magnetic field naturally points to cyclotron resonance as the origin of the coherent response, and to confirm this, we Fourier transform (see Methods) the oscillatory component of $r(t)$ obtained at different values of magnetic field and density. Selected line cuts at intervals of 1 T are shown in Fig. 4c (left axis). Consistent with cyclotron resonance, the oscillation frequency is found to depend linearly on the applied magnetic field for $B > 4$ T (Fig. 4c, right axis), allowing us to extract an effective mass $m^* \approx (0.328\pm0.008)m_e$ (see Methods). This value of $m^*$ is in reasonable agreement with reported measurements of $m^*$, including Shubnikov-de Haas analyses, which have yielded estimates ranging from $0.35m_e$ to $0.45m_e$[32–34], and with direct THz transmission measurement on monolayer WSe$_2$, which yielded $(0.5\pm0.09)m_e$[7]. We note that while Kohn's theorem[40] implies that the cyclotron mass cannot be renormalized by interactions in a spin-unpolarized Γ-valley system, the situation in WSe$_2$ is likely complex due to its $K$-valley Fermi surface and spin polarization at high magnetic field. Comparison with Fig. 3c raises the possibility that in the vicinity of the band edge, the cyclotron mass is not subject to possible interaction-induced renormalizations that govern plasmon propagation.

To identify the origin of the coherent exciton-cyclotron oscillations, we note that prior studies of the effects of moderate-strength THz fields on the order of kV/cm on Landau-quantized GaAs quantum wells have reported evidence for electronic redistribution across several Landau levels in the vicinity of the Fermi level, yielding time-oscillating level populations detected as re-emission at $\omega_c$[41,42]. To understand the effect of the THz field on the polaron resonances investigated here, we note that the THz field induces time-dependent broadenings $\Gamma(t)$ to the polaron resonances (Extended Data Figs. 3 and 4). The broadenings $\Gamma(t)$ are in turn controlled to lowest order by the occupation of the $n$th Landau level $\rho_n(t)$ via a "golden rule"-like expression $\Gamma \propto \sum_n \rho_n (1 - \rho_n)$ [38,39]. Under the driving field, the $\rho_n(t)$ oscillate at $\omega_c$ (see Methods and Supplementary Information), yielding an overall polaron line broadening oscillating at $\omega_c$. Examination of the reflectance contrast line shapes in the presence of the drives indicates complex behavior, likely including broadening in addition to potential redshifts and variations in oscillator strength (Extended Data Fig. 4). Consistent with this picture, we estimate field strengths in the range of several kV/cm, corresponding to effective Rabi frequencies exceeding the estimated Landau level widths (see Methods). We note that propagation-induced absorption (analogous to the mechanism by which plasmons are detected) can be ruled out due to the lack of an available propagating mode at high magnetic fields—indeed, the coherent oscillations do not exhibit substantial position dependence in the vicinity of the injection point (see Supplementary Information)—as well as an absence of oscillatory response in THz transmission measurements (see Methods). Comparison with the calculated quantum dynamics of the exciton resonance yields satisfactory agreement with the experiment (Fig. 4d), and thus demonstrates a second mechanism of THz-polaron coupling: rather than modifying the local electron density as in the $B=0$ case, at high magnetic field the pulse alters the energetic distribution of the electrons, which in turn modifies the polaron resonance due to its dependence on the Fermi surface. Overall, we conclude that our observations, which are inaccessible in THz transmission, enable us to access coherent quantum dynamics in an on-chip THz setting, and offer a promising means of probing meV-scale excitation gaps in vdW systems.

**Conclusion and outlook**

Our results establish a new approach for characterizing the THz electrodynamics of vdW materials that circumvents challenges associated with impedance matching in conventional on-chip THz transmission schemes. The coupling between THz excitations and optical polarons opens the door to studying the meV-scale excitation gaps and plasmonic collective modes in TMD moiré superlattices. Finally, our work provides a new solid-state platform for studying the dynamics of polarons[43] driven by electromagnetic fields[44,45], a regime not typically accessible in ultracold gases realizing electrically neutral driven polarons[46–50].

**References**

1. Andrei, E. Y. *et al.* The marvels of moiré materials. *Nat. Rev. Mater.* **6**, 201–206 (2021).

2. Gallagher, P. *et al.* Quantum-critical conductivity of the Dirac fluid in graphene. *Science* **364**, 158–162 (2019).

3. Island, J. O. *et al.* On-chip terahertz modulation and emission with integrated graphene junctions. *Appl. Phys. Lett.* **116**, 161104 (2020).

4. Seo, J. *et al.* On-Chip Terahertz Spectroscopy for Dual-Gated van der Waals Heterostructures at Cryogenic Temperatures. *Nano Lett.* **24**, 15060–15067 (2024).

5. Kipp, G. *et al.* Cavity electrodynamics of van der Waals heterostructures. *Nat. Phys.* (2025) doi:10.1038/s41567-025-03064-8.

6. Yoshioka, K., Kumada, N., Muraki, K. & Hashisaka, M. On-chip coherent frequency-domain THz spectroscopy for electrical transport. *Appl. Phys. Lett.* **117**, 161103 (2020).

7. Chen, S.-D. *et al.* Direct measurement of terahertz conductivity in a gated monolayer semiconductor. *Nano Lett.* **25**, 7998–8002 (2025).

8. Chen, S.-D. *et al.* Terahertz electrodynamics in a zero-field Wigner crystal. Preprint at https://doi.org/10.48550/arXiv.2509.10624 (2025).

9. Potts, A. M. *et al.* On-Chip Time-Domain Terahertz Spectroscopy of Superconducting Films below the Diffraction Limit. *Nano Lett.* **23**, 3835–3841 (2023).

10. Subedi, S. *et al.* Colossal terahertz emission with ultrafast tunability based on van der Waals ferroelectric $NbOI_2$. *Adv. Opt. Mater.* **13**, 2403471 (2025).

11. Kumar, R. K. *et al.* Terahertz photocurrent probe of quantum geometry and interactions in magic-angle twisted bilayer graphene. *Nat. Mater.* **24**, 1034–1041 (2025).

12. Bandurin, D. A. *et al.* Cyclotron resonance overtones and near-field magnetoabsorption via terahertz Bernstein modes in graphene. *Nat. Phys.* **18**, 462–467 (2022).

13. Venanzi, T. *et al.* Terahertz-Induced Energy Transfer from Hot Carriers to Trions in a MoSe2 Monolayer. *ACS Photonics* **8**, 2931–2939 (2021).

14. Venanzi, T. *et al.* Ultrafast switching of trions in 2D materials by terahertz photons. *Nat. Photonics* **18**, 1344–1349 (2024).

15. Shi, J. *et al.* Room Temperature Terahertz Electroabsorption Modulation by Excitons in Monolayer Transition Metal Dichalcogenides. *Nano Lett.* **20**, 5214–5220 (2020).

16. Hoegen, A. von *et al.* Visualizing a Terahertz Superfluid Plasmon in a Two-Dimensional Superconductor. Preprint at https://doi.org/10.48550/arXiv.2506.08204 (2025).

17. Mak, K. F. *et al.* Tightly bound trions in monolayer MoS2. *Nat. Mater.* **12**, 207–211 (2013).

18. Ross, J. S. *et al.* Electrical control of neutral and charged excitons in a monolayer semiconductor. *Nat. Commun.* **4**, 1474 (2013).

19. Sidler, M. *et al.* Fermi polaron-polaritons in charge-tunable atomically thin semiconductors. *Nat. Phys.* **13**, 255–261 (2017).

20. Michael, M. H. *et al.* Resolving self-cavity effects in two-dimensional quantum materials. Preprint at https://doi.org/10.48550/arXiv.2505.12799 (2025).

21. Fei, Z. *et al.* Gate-tuning of graphene plasmons revealed by infrared nano-imaging. *Nature* **487**, 82–85 (2012).

22. Chen, J. *et al.* Optical nano-imaging of gate-tunable graphene plasmons. *Nature* **487**, 77–81 (2012).

23. Tulyagankhodjaev, J. A. *et al.* Room-temperature wavelike exciton transport in a van der Waals superatomic semiconductor. *Science* **382**, 438–442 (2023).

24. Bae, Y. J. *et al.* Exciton-coupled coherent magnons in a 2D semiconductor. *Nature* **609**, 282–286 (2022).

25. Low, T. *et al.* Polaritons in layered two-dimensional materials. *Nat. Mater.* **16**, 182–194 (2017).

26. Ju, L. *et al.* Graphene plasmonics for tunable terahertz metamaterials. *Nat. Nanotechnol.* **6**, 630–634 (2011).

27. Koppens, F. H. L., Chang, D. E. & García de Abajo, F. J. Graphene Plasmonics: A Platform for Strong Light–Matter Interactions. *Nano Lett.* **11**, 3370–3377 (2011).

28. Jablan, M., Buljan, H. & Soljačić, M. Plasmonics in graphene at infrared frequencies. *Phys. Rev. B* **80**, 245435 (2009).

29. Eguiluz, A., Lee, T. K., Quinn, J. J. & Chiu, K. W. Interface excitations in metal-insulator-semiconductor structures. *Phys. Rev. B* **11**, 4989–4993 (1975).

30. Das Sarma, S. & Madhukar, A. Collective modes of spatially separated, two-component, two-dimensional plasma in solids. *Phys. Rev. B* **23**, 805–815 (1981).

31. Larentis, S. *et al.* Large effective mass and interaction-enhanced Zeeman splitting of $K$-valley electrons in MoSe2. *Phys. Rev. B* **97**, 201407 (2018).

32. Fallahazad, B. *et al.* Shubnikov--de Haas Oscillations of High-Mobility Holes in Monolayer and Bilayer WSe2: Landau Level Degeneracy, Effective Mass, and Negative Compressibility. *Phys. Rev. Lett.* **116**, 086601 (2016).

33. Joe, A. Y. *et al.* Transport Study of Charge-Carrier Scattering in Monolayer WSe2. *Phys. Rev. Lett.* **132**, 056303 (2024).

34. Pack, J. *et al.* Charge-transfer contacts for the measurement of correlated states in high-mobility WSe2. *Nat. Nanotechnol.* **19**, 948–954 (2024).

35. Zhang, Y. & Das Sarma, S. Quasiparticle effective-mass divergence in two-dimensional electron systems. *Phys. Rev. B* **71**, 045322 (2005).

36. Kwon, Y., Ceperley, D. M. & Martin, R. M. Quantum Monte Carlo calculation of the Fermi-liquid parameters in the two-dimensional electron gas. *Phys. Rev. B* **50**, 1684–1694 (1994).

37. Sun, Z., Basov, D. N. & Fogler, M. M. Universal linear and nonlinear electrodynamics of a Dirac fluid. *Proc. Natl. Acad. Sci.* **115**, 3285–3289 (2018).

38. Smoleński, T. *et al.* Interaction-Induced Shubnikov--de Haas Oscillations in Optical Conductivity of Monolayer MoSe2. *Phys. Rev. Lett.* **123**, 097403 (2019).

39. Efimkin, D. K. & MacDonald, A. H. Exciton-polarons in doped semiconductors in a strong magnetic field. *Phys. Rev. B* **97**, 235432 (2018).

40. Kohn, W. Cyclotron Resonance and de Haas-van Alphen Oscillations of an Interacting Electron Gas. *Phys. Rev.* **123**, 1242–1244 (1961).

41. Arikawa, T. *et al.* Quantum control of a Landau-quantized two-dimensional electron gas in a GaAs quantum well using coherent terahertz pulses. *Phys. Rev. B* **84**, 241307 (2011).

42. Maag, T. *et al.* Coherent cyclotron motion beyond Kohn's theorem. *Nat. Phys.* **12**, 119–123 (2016).

43. Massignan, P. *et al.* Polarons in atomic gases and two-dimensional semiconductors. Preprint at https://doi.org/10.48550/arXiv.2501.09618 (2025).

44. Van Tuan, D. Marrying Excitons and Plasmons in Monolayer Transition-Metal Dichalcogenides. *Phys. Rev. X* **7**, (2017).

45. Van Tuan, D. Probing many-body interactions in monolayer transition-metal dichalcogenides. *Phys. Rev. B* **99**, (2019).

46. Kohstall, C. *et al.* Metastability and coherence of repulsive polarons in a strongly interacting Fermi mixture. *Nature* **485**, 615–618 (2012).

47. Adlong, H. S. Quasiparticle Lifetime of the Repulsive Fermi Polaron. *Phys. Rev. Lett.* **125**, (2020).

48. Vivanco, F. J. *et al.* The strongly driven Fermi polaron. *Nat. Phys.* **21**, 564–569 (2025).

49. Wasak, T. Decoherence and Momentum Relaxation in Fermi-Polaron Rabi Dynamics: A Kinetic Equation Approach. *Phys. Rev. Lett.* **132**, (2024).

50. Parish, M. M. Quantum dynamics of impurities coupled to a Fermi sea. *Phys. Rev. B* **94**, (2016).

51. Garg, R., Bahl, I. J. & Bozzi, M. *Microstrip Lines and Slotlines*. (Artech House, Boston, 2013).

52. Grischkowsky, D., Keiding, S., van Exter, M. & Fattinger, Ch. Far-infrared time-domain spectroscopy with terahertz beams of dielectrics and semiconductors. *J. Opt. Soc. Am. B* **7**, 2006 (1990).

53. Staffaroni, M., Conway, J., Vedantam, S., Tang, J. & Yablonovitch, E. Circuit analysis in metal-optics. *Photonics Nanostructures - Fundam. Appl.* **10**, 166–176 (2012).

54. Oskooi, A. F. *et al.* Meep: A flexible free-software package for electromagnetic simulations by the FDTD method. *Comput. Phys. Commun.* **181**, 687–702 (2010).

55. Efimkin, D. K. & MacDonald, A. H. Many-body theory of trion absorption features in two-dimensional semiconductors. *Phys. Rev. B* **95**, 035417 (2017).

56. Efimkin, D. K., Laird, E. K., Levinsen, J., Parish, M. M. & MacDonald, A. H. Electron-exciton interactions in the exciton-polaron problem. *Phys. Rev. B* **103**, 075417 (2021).

57. Hagenmüller, D., De Liberato, S. & Ciuti, C. Ultrastrong coupling between a cavity resonator and the cyclotron transition of a two-dimensional electron gas in the case of an integer filling factor. *Phys. Rev. B* **81**, 235303 (2010).

**Methods**

**Device fabrication and characterization.** Transmission lines consisting of 10 nm Ti and 50 nm Au are fabricated on an insulating Si wafer with 285 nm dry thermal oxide via photolithography and electron beam evaporation. vdW heterostructures are fabricated using polymer stamps consisting of poly(bisphenol A carbonate), poly(propylene carbonate) and polydimethylsiloxane; the heterostructures consist of graphite-contacted monolayer $WSe_2$ and $MoSe_2$ (HQGraphene) encapsulated by hexagonal boron nitride of typical thickness 40 nm. The vdW stacks are placed down directly onto the electrodes. Photoconductive switches are fabricated from low-temperature-grown GaAs (BATOP GmbH) subjected to 10 min annealing at 600 C in a rapid thermal annealer, contacted with Cr/Au following a hydrofluoric acid dip and released using a standard epitaxial release process[2]. The switches are placed onto the waveguide via the same dry transfer technique used for stacking the vdW heterostructures. The vdW device is shown in Extended Data Fig. 5.

**THz-pump optical-probe measurement scheme**. Laser pulses of duration approximately 100 fs at 1030 nm (Light Conversion Pharos) and repetition rate 1 MHz are diverted into two separate beam paths to perform the THz pump-optical probe measurements. First, a portion of the 1030 nm beam is frequency-doubled to 515 nm, approximately 0.2 to 1 mW of which is used to excite the photoconductive switch and generate pulses on the transmission line. The remainder of the 1030 nm beam is directed to a YAG crystal for broadband supercontinuum generation. Approximately 1 µW of the supercontinuum pulse, obtained after spectral filtering to cover the $WSe_2$ 1s resonance, is directed to the device for optical spectroscopy. The device is situated in a vibration-isolated cryostat with a base temperature of 3.5 K equipped with a 9 T magnet (AttoDry 1000). The time delay between the pump and probe beams is controlled via a mechanical delay stage. The reflected portion of the probe beam is spectrally dispersed in a grating spectrometer and read out with a CCD camera. The probe and pump beams are chopped at 1 kHz and 500 Hz, respectively, with the camera synchronized to the choppers for high-fidelity readout of $\Delta R/R$. The zero-delay time is determined from the peak of $r(t)$ measured at $n_{hole}=3.8\times10^{12}$ $cm^{-2}$ where the observed optical response is fastest.

**THz transmission measurements.** Laser pulses of duration approximately 100 fs at 800 nm generated by an 80 MHz Ti:sapphire oscillator are diverted into two separate beam paths as in the measurement protocol described above. One beam path is used for generating THz pulses as described above. Pulses from the second beam path are diverted to a second photoconductive switch situated on the transmission line intended to serve as a detector in a standard scheme[2–9].

**Transmission line characteristics.** In the simplest approximation, the characteristic impedance $Z_0^{CPS}$ of the quasi-TEM mode of the coplanar strip transmission line used to deliver THz pulses to the vdW device is given by[51]

$$Z_0^{CPS} = \sqrt{\frac{\mu_0}{\varepsilon_{\text{eff.}}}} \frac{K(W/W_{\text{tot.}})}{K[\sqrt{1-(W/W_{\text{tot.}})^2}]}$$

where $K$ is the complete elliptic integral of the first kind, $W$ is the distance between the transmission line electrodes, $W_{\text{tot.}}$ is the width of the entire structure including that of the electrodes (i.e., $W_{\text{tot.}} = W + 2W_{\text{electrode}}$, where $W_{\text{electrode}}$ is the electrode width), and the effective dielectric constant $\varepsilon_{\text{eff.}}$ is given by the average of that of vacuum and that of the Si substrate, i.e. $\varepsilon_{\text{eff.}} = \varepsilon_0(1+\varepsilon_{r,\text{Si}})/2$, where $\varepsilon_{r,\text{Si}}$ is approximately 11.6 in the frequency range of interest[52]. For the geometry used in our devices, $W$ is 15 μm and $W_{\text{tot.}}$ is 30 μm, yield $Z_0 \approx 117$ Ω. The propagation velocity $v_{\text{eff.}}$ of this mode is given approximately by

$$v_{\text{eff.}} = \frac{c}{\sqrt{(1+\varepsilon_{r,\text{Si}})/2}} \approx c/2.51.$$

**Extraction of $r(t)$ line traces from full $\Delta R/R$ spectra.** $r(t)$ traces are determined from full spectra $\Delta R(E, t)/R$ via the relation

$$r(t) = \frac{\int_{E_-}^{E_+} dE \left|\frac{\Delta R(E,t)}{R}\right|}{\max_t \left(\int_{E_-}^{E_+} dE \left|\frac{\Delta R(E,t)}{R}\right|\right)},$$

i.e. by integrating the absolute value of $\Delta R/R$ with respect to photon energy over the range spanning from $E_-$ and $E_+$, which define the photon energy range corresponding to the repulsive polaron resonance, and subsequently normalizing by the value at which the time at which the maximum is achieved (denoted $\max_t$).

**Propagation velocity extraction.** Velocities are extracted by measuring $r(t)$ in the vicinity of the injection electrode, where the effects of propagation-induced broadening are the smallest and the peak of the propagating pulse translates approximately linearly in time. Measurements are performed at four spatial locations over the full range of densities shown. Raw times-of-flight are determined by fitting the peaks of $r(t)$ with a polynomial and are then converted to distances by comparing the initial and final beam spot positions obtained from CMOS images. The uncertainty in the position determination is taken to be ±1 μm, set by the beam spot as well as the precision with which the images can be aligned. Finally, velocities are obtained by performing orthogonal distance regressions on the time-versus-position data obtained from the above procedure, where the position uncertainty is included, as well as the total travel time uncertainty set by the delay stage resolution and the standard-deviation polynomial fit uncertainty. The error bars in Fig. 3d reflect the standard deviation of the orthogonal distance regression.

**Effective mass extraction from time-domain cyclotron resonance.** To extract effective masses from the measured time-domain cyclotron resonance, we extract $r(t)$ at a given density and magnetic field and Fourier transform it using the fast Fourier transform (FFT) algorithm. The raw FFTs are Hann windowed, and a smooth background corresponding to the frequency rolloff of the input pulse is subtracted. The peaks corresponding to $\omega_c$ as in Fig. 4c are extracted and fit via the relation $f_c=\omega_c/2\pi=eB/2\pi m^*$, where $f_c$ is the cyclotron resonance frequency. We note that, despite measuring down to carrier densities of approximately $2\times10^{12}$ cm$^{-2}$, the measured cyclotron oscillations give no indication of a reported suppression of the effective mass related to spin polarization[34].

**Density axis calibration.** To determine the device capacitance, we measure $\Delta R/R$ at fixed delay and as a function of magnetic field, resulting in a Landau fan-like pattern (Extended Data Fig. 6). The charge neutrality point is determined by identifying the gate voltage at which the $\Delta R/R$ response vanishes from a separate fixed-delay gate-dependent measurement at zero magnetic field.

**Plasmon modes in layered media.** In general, surface plasmons arising from interfacial conductors take the form of transverse-magnetic (TM) modes satisfying $\mathbf{H} = H_y\hat{y}$ and $\mathbf{E} = E_x\hat{x} + E_z\hat{z}$ in terms of the coordinate system specified in Fig. 1. Assuming dependence on $z$ and $t$ of the form $e^{iqz-i\omega t}$ gives for a single conducting plane a complex wavevector $q=2i\omega\varepsilon_{\text{eff}}/\sigma(\omega)$ at sufficiently high frequencies[21,28,53]. Within a Drude model description, $\sigma(\omega)$ in turn takes the form $\sigma(\omega)=(R_s-i\omega L_k)^{-1}$, where $L_k=m^*/ne^2$ is the kinetic inductance and $R_s=L_k/\tau_{\text{Drude}}$ is the sheet resistance. For the case of WSe$_2$, at the density $n_{\text{hole}} \approx 4.1\times10^{11}$ cm$^{-2}$ studied above and with a width of 15 μm, the expected kinetic inductance per unit length $L_k/W \sim m^*/ne^2W$ is approximately $3\times10^{-4}$ H/m, exceeding by more than two orders of magnitude the geometric contribution on the order of $\mu_0 \approx 2\times10^{-6}$ H/m.[53] For a general multilayered structure, the above ansatz leads to a matrix equation relating the amplitudes $H_y^{i\pm}$ in each layered medium, where the index $i$ specifies the medium and ± the two independent field components within the medium. For the case of a four-layer structure, the matrix equation can be conveniently reduced to a univariate equation of form $F(q) = 0$, where $F$ is a transcendental function, that is straightforward to solve numerically for all modes $q(\omega)$. Full details of the derivation and the form of $F$ are given in the Supplementary Information. Group velocities are calculated via the relation $v_{\text{group}}=d\omega/dq$.

**Electromagnetic simulations.** We perform finite-difference time domain calculations using the MEEP package[54]. We assume translation invariance in the $y$ direction and study a system with the geometry shown in the inset to Fig. 3a. Excitation of the optical plasmon mode is conveniently achieved via a magnetic charge-current source directed in the $y$ direction, which results in circulating electric fields producing the required $x$ and $z$ components of the mode. For computational convenience, the resolution is fixed at 40 nm and the geometry is truncated after ±10 microns in the $z$ direction. Examples of the calculated field transients are shown in Extended Data Fig. 7. A propagation velocity is extracted in each case by determining the arrival time of the pulse center (determined by the zero crossing of the electric field) at each spatial position between 2 and 7 μm. Increasing the resolution to 20 nm is found to change the calculation results negligibly. The Drude scattering time $\tau$ is taken to be 2 ps.

**Quantum mechanical model of exciton-coupled coherent cyclotron response.** Full details of the quantum-mechanical calculation describing the effect of the $\omega_c$-resonant drive on the polaron linewidths are given in the Supplementary Information. Briefly, within a Fermi polaron picture[19,39,55,56] we consider the many-body Hamiltonian

$$H = \sum_{nn',\alpha\alpha'} c_{n\alpha}^\dagger c_{n'\alpha'} \langle n\alpha|H_0 + V|n'\alpha'\rangle + \sum_k E_k d_k^\dagger d_k - \frac{U}{S}\sum_{kp}\sum_{nn',\alpha\alpha'} d_k^\dagger d_p c_{n'\alpha'}^\dagger c_{n\alpha} F_{nn'}^{\alpha\alpha'}(k-p)$$

where $c_{n\alpha}$ and $d_k$ are respectively the annihilation operators of electrons in orbital $\alpha$ of the $n$th Landau level and excitons at momentum $k$, $H_0$ is the single-electron kinetic energy, $V \propto A_x(t) - iA_y(t)$ is the contribution of the applied time-dependent electric field (in terms of the vector potential $A$), $E_k$ is the exciton energy, $U$ is an attractive coupling originating from a charge-dipole interaction, S is the system area and $F_{nn'}^{\alpha\alpha'}$ is a form factor (see Supplementary Information). Standard diagrammatic perturbation theory enables computation of the energy-dependent self-energy $\Sigma_0$ at order $U^2$, which in turn yields the electronic contribution to the broadening of the exciton line $\Gamma$ via $\Gamma = -\text{Im}(\Sigma_0)$, with $\Sigma_0$ evaluated at the polaron energy. At zero temperature, $\Gamma$ is determined by the time-dependent $n$th Landau level populations $\rho_n(t)$ via[38]

$$\Gamma(t) = \frac{m_X \pi U^2 eB}{\hbar^3} \sum_n \rho_n(t)[1-\rho_n(t)],$$

where $m_X$ is the exciton mass, consistent with the observation that the field-induced electron redistribution modifies the line broadening; inter-Landau-level scattering, which is not contained in the equation above, mostly modifies the high-energy tails of the polaronic peaks. To obtain the $\rho_n(t)$, we solve the Heisenberg equation of motion for the operators $c$, which to lowest order in $V$ are found to exhibit time dependence proportional to

$$\delta\rho_n(t) \propto \left|\int_0^t dt'\, V(t') e^{-i\omega_c t'}\right|^2,$$

consistent with the observation that excitation by a broadband pulse $V(t)$ yields a time-dependent linewidth oscillating at $\omega_c$ (Extended Data Fig. 3). We note that the rate derived perturbatively in the exciton-electron coupling cannot fully describe the attractive polaron. The latter requires a non-perturbative evaluation of the two-particle $T$-matrix, which is challenging at finite magnetic fields. However, qualitatively we expect that the linewidth of the attractive polaron is determined by the Landau level population as well. We note that, in the experiment, prominent oscillations are observed in both the RP and AP branches. Finally, we note that in a given device, the measured optical spectrum represents the real part of the exciton propagator, leading to sign changes observed experimentally (see Supplementary Information).

**THz field amplitude estimation.** To estimate the peak field strength, we first define an effective duty cycle via

$$d_{\text{eff.}} = f_{\text{rr}} \int_0^{t_{\text{pulse}}} I_{\text{norm.}}(t)\, dt$$

where $f_{\text{rr}}$ is the laser repetition rate, $I_{\text{norm.}}(t)$ is the THz current transient measured using the scheme above normalized to a peak value of 1 and $t_{\text{pulse}}$ is the pulse duration (indicated with the right-hand grey line in Extended Data Fig. 1). The transient amplitude is estimated from the time-average DC current $I_{\text{DC}}$ flowing along the transmission line, which we monitor in our time-resolved reflectance measurements. In terms of $d_{\text{eff.}}$, the peak electric field $E_{\text{peak}}$ is approximately given by

$$E_{\text{peak}} \approx \frac{1}{2W}(2I_{\text{DC}})Z_0 d_{\text{eff.}}^{-1} = \frac{I_{\text{DC}}Z_0}{W d_{\text{eff.}}}$$

where the factor of 2 accounts for the duty cycle of the chopper and the factor of ½ accounts for the fact that pulses are injected into the transmission line in two directions simultaneously, only one of which is effective in exciting the sample. The experiments discussed in the text use values of $I_{\text{DC}}$ ranging from approximately 50 to 260 nA, which we estimate to correspond to approximately 1.3 to 6.9 kV/cm.

**Cyclotron Rabi frequency at high magnetic field.** The effective Rabi frequency associated with driven cyclotron transitions is approximately given by $\hbar\Omega_{\text{Rabi}} = E_{\text{THz}} \times el_B\sqrt{\nu}$, where $l_B = \sqrt{\hbar/eB}$ is the magnetic length and $\nu$ is the Landau level filling factor.[57] Making use of the photocurrent-based method above, with a DC photocurrent of 260 nA, we estimate the peak value of $E_{\text{THz}}$ at 9 T to be approximately 6.9 kV/cm. At the density of $n=4.1\times10^{11}$ cm$^{-2}$ and at 9 T, the filling factor is approximately 1.8, and we find a Rabi frequency of approximately $\Omega_{\text{Rabi}}/2\pi=1.9$ THz. This represents a lower bound on the Rabi frequency over the parameter space measured in the experiment, as reducing the magnetic field and increasing the density will both increase the Rabi frequency. Thus, the Rabi frequency is comparable to or exceeds the Landau level width (inferred from the apparent oscillations lifetimes of several ps), as required for repopulation effects to occur. We note that the Rabi frequency is not reflected directly in the oscillation frequency for a harmonic oscillator as it would be for a two-level system; instead, the frequency of the oscillations is always $\omega_c$ (see Supplementary Information).

**Data availability**

The data that supports the findings of this study are available from the corresponding authors upon reasonable request.

**Code availability**

The codes that supports the findings of this study are available from the corresponding authors upon reasonable request.

**Acknowledgments**

We acknowledge discussions with Ben Feldman. This work was supported by the US Department of Energy, Office of Science, Basic Energy Sciences, under award number DE-SC0019481 (measurements), the National Science Foundation DMR-1807810 (sample fabrication) and the Air Force Office of Scientific Research under award number FA9550-20-1-0219 (modelling). It was also funded in part by the Gordon and Betty Moore Foundation (grant number GBMF11563). We used the Cornell NanoScale Facility, an NNCI member supported by NSF Grant NNCI-2025233, for sample fabrication. E. M. acknowledges support from the National Science Foundation under grant number NSF PHY-2409403. K.W. and T.T. acknowledge support from the JSPS KAKENHI (Grant Numbers 21H05233 and 23H02052), the CREST (JPMJCR24A5), JST and World Premier International Research Center Initiative (WPI), MEXT, Japan. A.T.P. acknowledges support from the Kavli Institute at Cornell (KIC) Postdoctoral Fellowship. D.P. was supported by the German Research Foundation (DFG) under Project-ID 442134789.

**Author Contributions**

A.T.P., C.V., K.F.M. and J.S. designed the experiment. A.T.P. and C.V. performed the measurements with input from K.F.M. and J.S. A.T.P., C.V. and S. X. fabricated the devices. A.T.P and S.X. performed electromagnetic calculations. D.P., D.C. and E. M. developed the quantum mechanical model. K.W. and T.T. provided hBN crystals. All authors participated in discussions and in writing the manuscript.

**Competing interests**

The authors declare no competing interests.

**Figure 1**

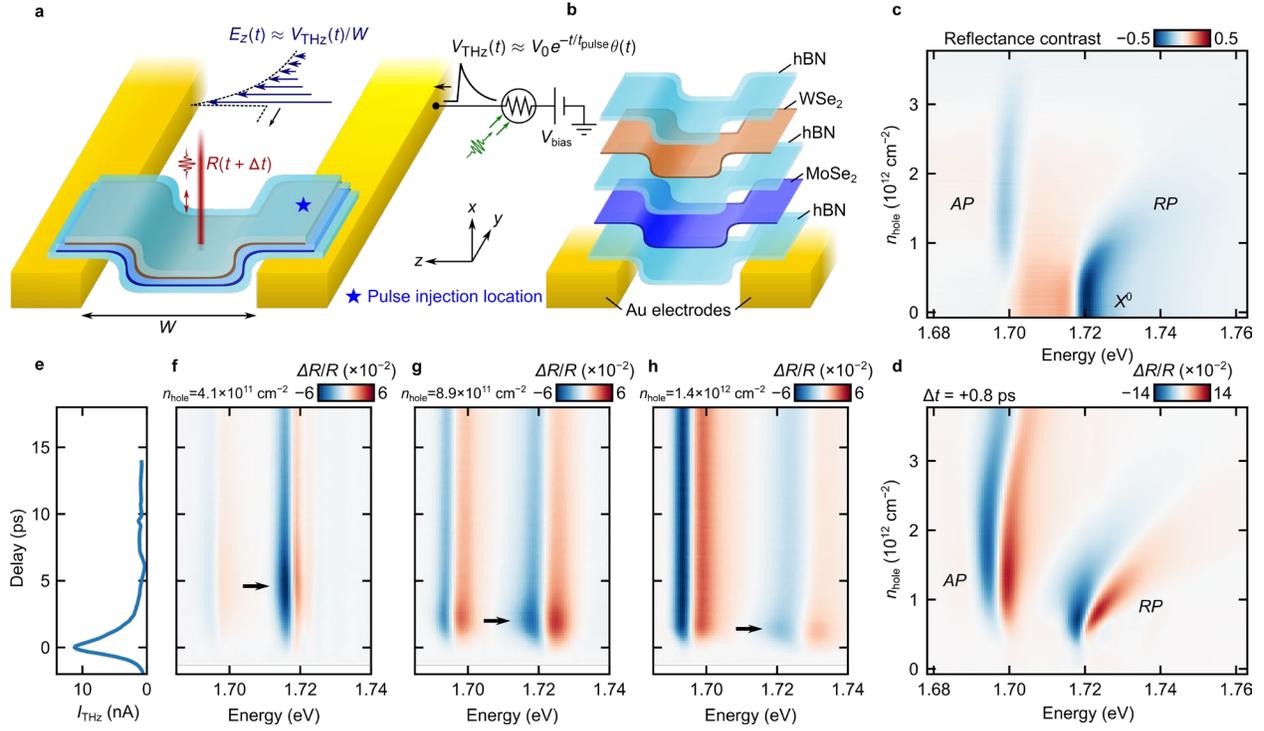

**Fig. 1. | Polaron sensing of nonequilibrium THz dynamics in vdW heterostructures. a, b,** schematic diagrams depicting the experimental setup (a), with the vdW heterostructure embedded in a transmission line and subject to THz pumping and near infrared probing fields, and the layout of the vdW heterostructure (b) incorporating hBN-encapsulated $WSe_2$ and $MoSe_2$ monolayers. **c,** static reflectance spectrum of $WSe_2$ measured as a function of carrier density in the vicinity of the 1s exciton resonance ($T$=3.5K and $B$=0T). **d,** measurement of $\Delta R/R$ at a fixed delay of $t$=+0.8 ps after the arrival of the THz transient as a function of carrier density. **e,** picosecond-timescale voltage transient coupled to the transmission line used to drive the vdW heterostructure. **f-h,** measurements of $\Delta R(E, t)/R(E, t)$ spectra exhibiting response to the incident THz pulse. The *y* axes for panels f-h coincide with that of panel e.

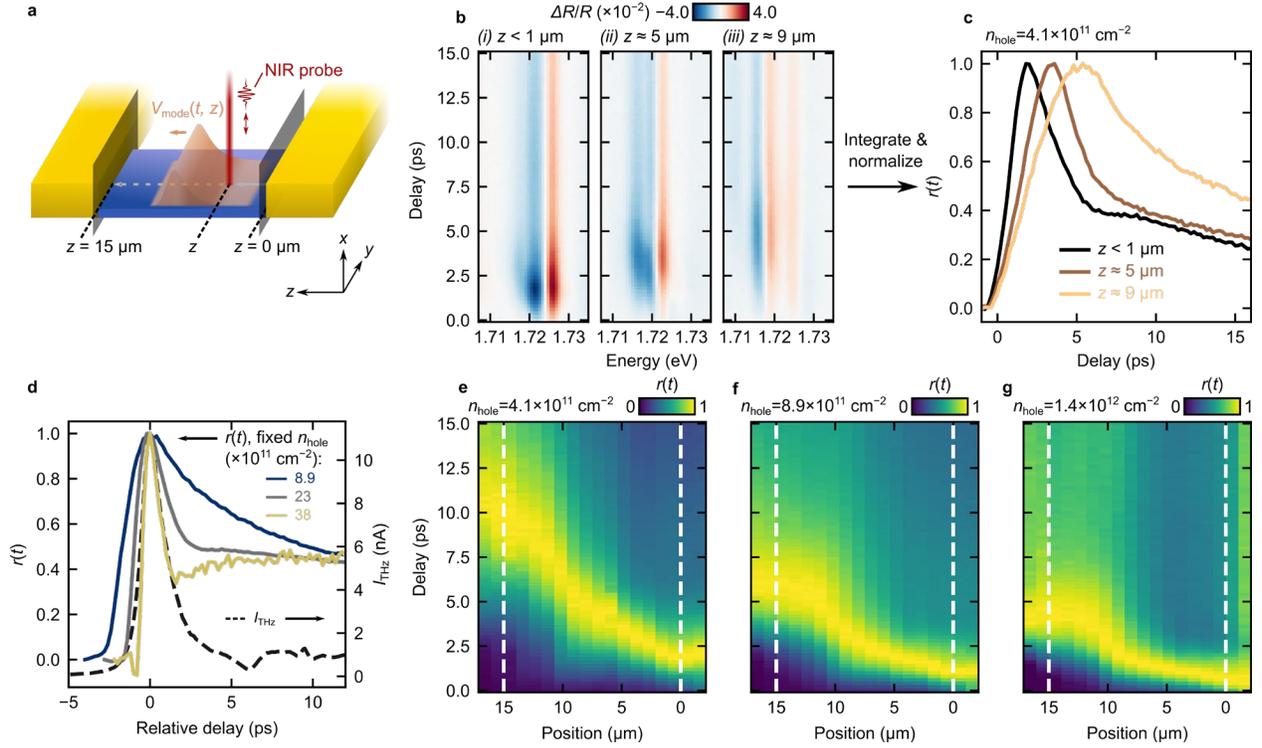

**Fig. 2 | Picosecond-timescale electrodynamic imaging. a,** sketch illustrating the geometry for position-dependent $\Delta R/R$ measurements (see text). The orange waveform represents a propagating plasmon mode that is sampled at position $z$ by the near-infrared probe beam. **b,** $\Delta R/R$ spectra measured as a function of delay time at different positions $z$ along the waveguide relative to the injection point at the right edge of the device (denoted $z$=0) and at a fixed density of $n_{hole}$=4.1×10$^{11}$ cm$^{-2}$ ($T$=3.5K and $B$=0T). **c,** example integrated r(t) transients obtained from the data in panel b. **d,** comparison of extracted $r(t)$ transients up to high hole doping overlaid on the independently measured THz voltage transient measured approximately 5 μm from the injection point. To illustrate the density-dependent broadening, the peak value of each trace is manually aligned to $t$=0. **e-g**, extracted values of the normalized quantity $r(t)$ as a function of position between the waveguide electrodes at densities of 4.1×10$^{11}$ cm$^{-2}$ (e), 8.9×10$^{11}$ cm$^{-2}$ (f) and 1.4×10$^{12}$ cm$^{-2}$ (g).

**Figure 3**

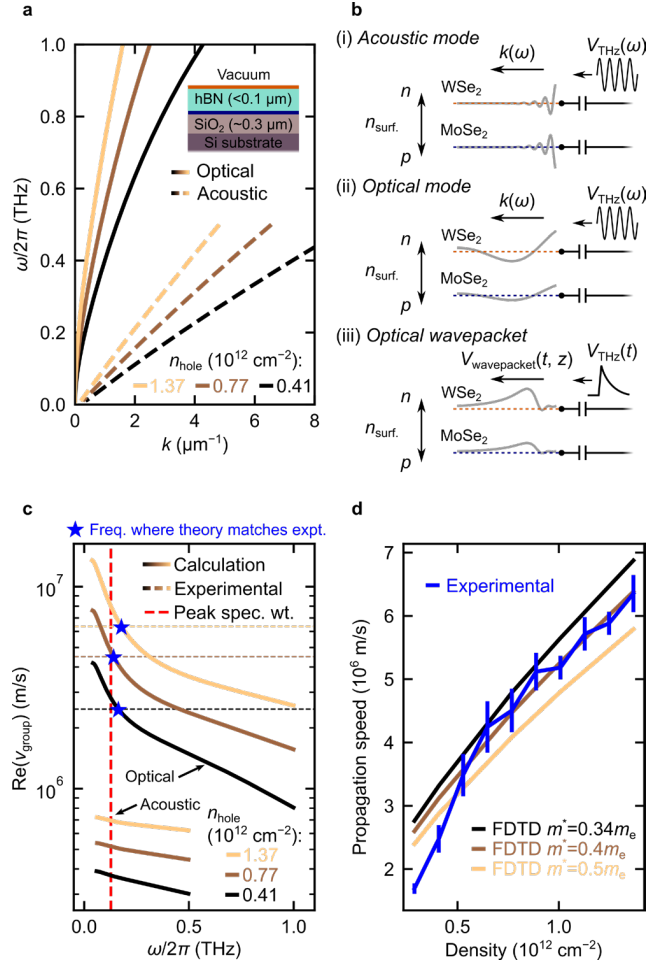

**Fig. 3. | Identifying the plasmon mode and propagation characteristics. a,** calculated dispersions of the optical and acoustic modes of the four-layer system (see text) at three representative densities. Inset: geometry used for analytic and numerical calculations. The orange and blue interfaces in the sketch are respectively the hole and electron layers. **b,** illustration of the oscillating charge configuration for the acoustic (i) and optical (ii) modes. The grey traces indicate the sign of the carriers on each layer with the dashed lines representing neutrality. Broadband excitation of the optical modes results in the injection of a wavepacket constructed from a superposition of optical modes (iii). **c,** group velocity of the acoustic and optical modes as a function of frequency. Experimental values are indicated with dash lines, and the frequency at which the experimental velocity intersects the calculated group velocity is marked with a star. The red dotted line indicates the estimated spectral weight center of the input pulse. **d,** comparison of the experimental density-dependent propagation speed with those obtained from FDTD calculations (see text). Error bars correspond to 1σ fit standard deviations (see Methods).

**Figure 4**

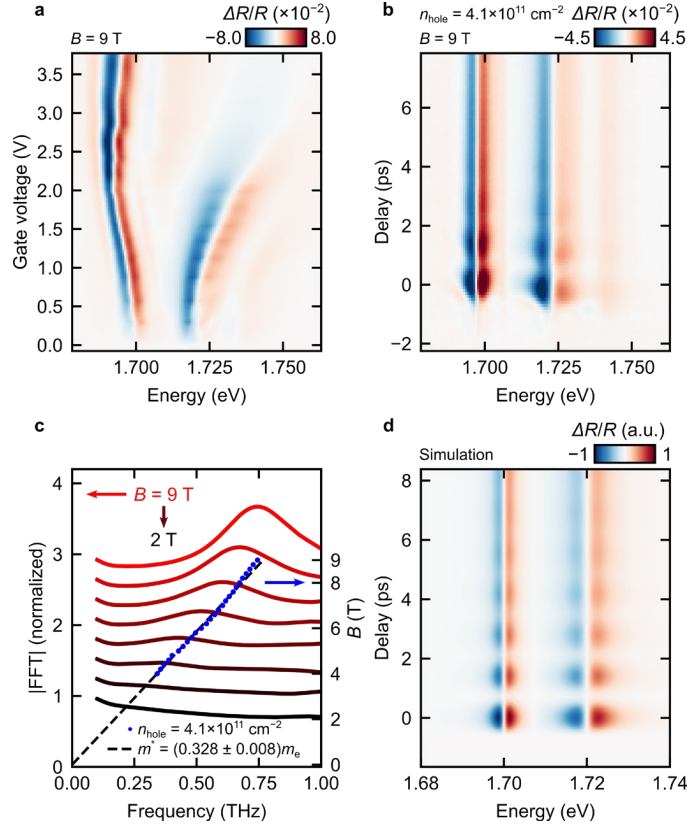

**Fig. 4. | Time-resolved quantum dynamics and cyclotron gap spectroscopy. a,** measurement of $\Delta R/R$ as a function of carrier density at a fixed delay ($t=-0.2$ ps) on the hole-doped side of WSe$_2$ at 9 T ($T=3.5$K). **b,** measurement of $\Delta R/R$ spectra exhibiting real-time cyclotron oscillations as a function of delay at 9 T and $n_{\text{hole}}$ equal to $4.1\times10^{11}$ cm$^{-2}$. **c,** left axis (red-black curves): Fourier transform of $r(t)$ extracted from measurements at a hole density of $n_{\text{hole}}=4.1\times10^{11}$ cm$^{-2}$ obtained as a function of magnetic field ($B=2$-9T) at a fixed spatial position. Curves are offset vertically for clarity and are normalized by the measured input pulse spectrum at zero carrier density. Right axis (blue points and dashed line): cyclotron energies $\hbar\omega_c$ as a function of magnetic field extracted from Fourier transforms obtained at $n_{\text{hole}}=4.1\times10^{11}$ cm$^{-2}$. Data points are overlaid on a best-fit line yielding $m^*=(0.328\pm0.008)m_e$. **d,** simulated time-dependent reflectance spectrum calculated using the quantum mechanical model (see text).

**Extended Data Figure 1**

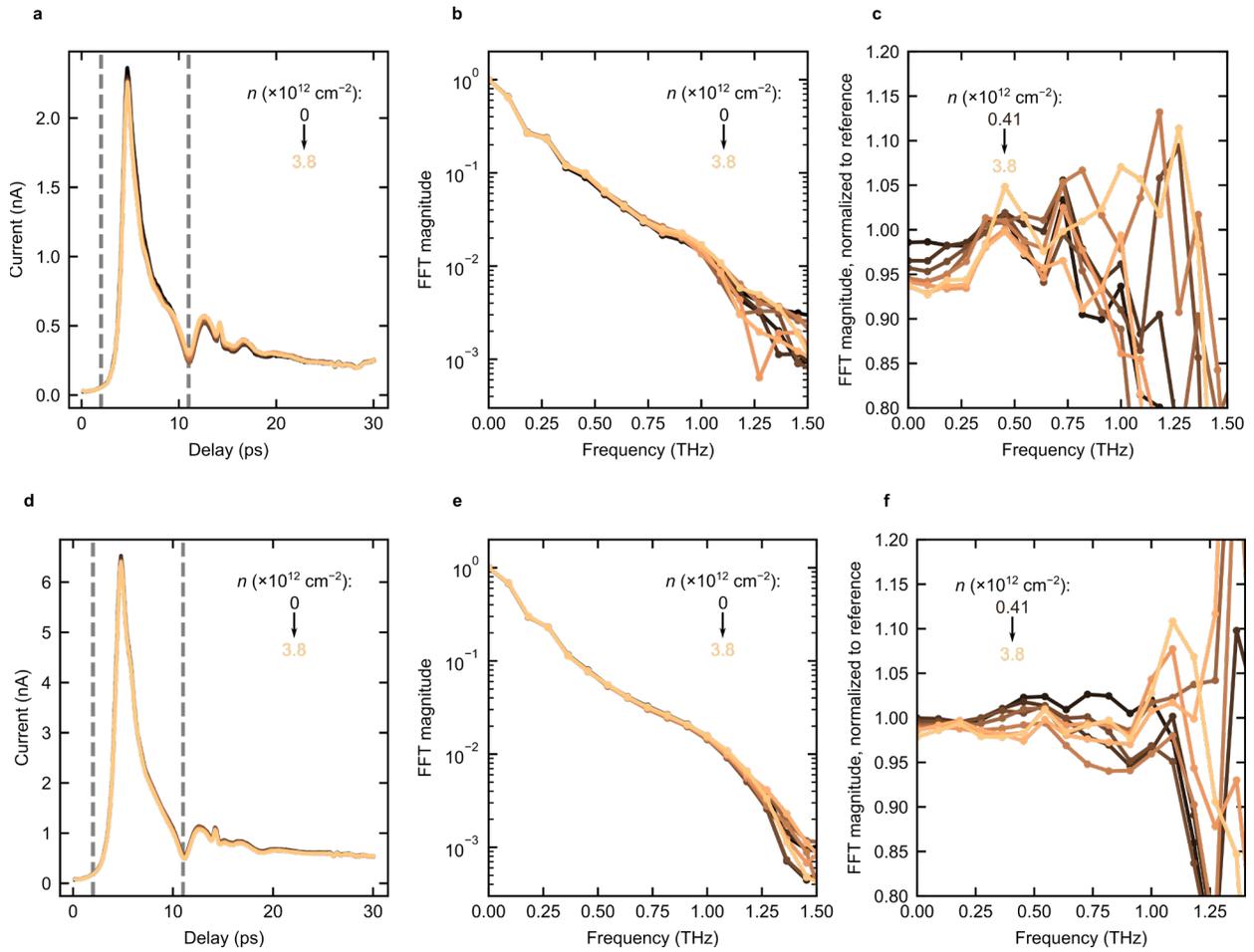

**Extended Data Fig. 1. | THz transmission measurements of the WSe$_2$/MoSe$_2$ device. a,** on-chip time-domain THz transmission transients recorded from densities of 0 to $3.8×10^{12}$ cm$^{−2}$ in steps of $4.8×10^{11}$ cm$^{−2}$. The grey dashed lines indicate the portion of the transient that is Fourier transformed. Features appearing after 10 ps are primarily reflections from distant locations on the transmission line. **b,** Fourier transform magnitudes derived from the time series indicated in panel a. **c,** Fourier transforms at finite density normalized by the reference spectrum obtained at charge neutrality. **d,** time-domain THz transients obtained at 9 T analogous to panel a. **e,** Fourier transform magnitudes derived from the time series indicated in panel d. **f,** 9 T Fourier transforms at finite density normalized by the reference spectrum obtained at charge neutrality.

**Extended Data Figure 2**

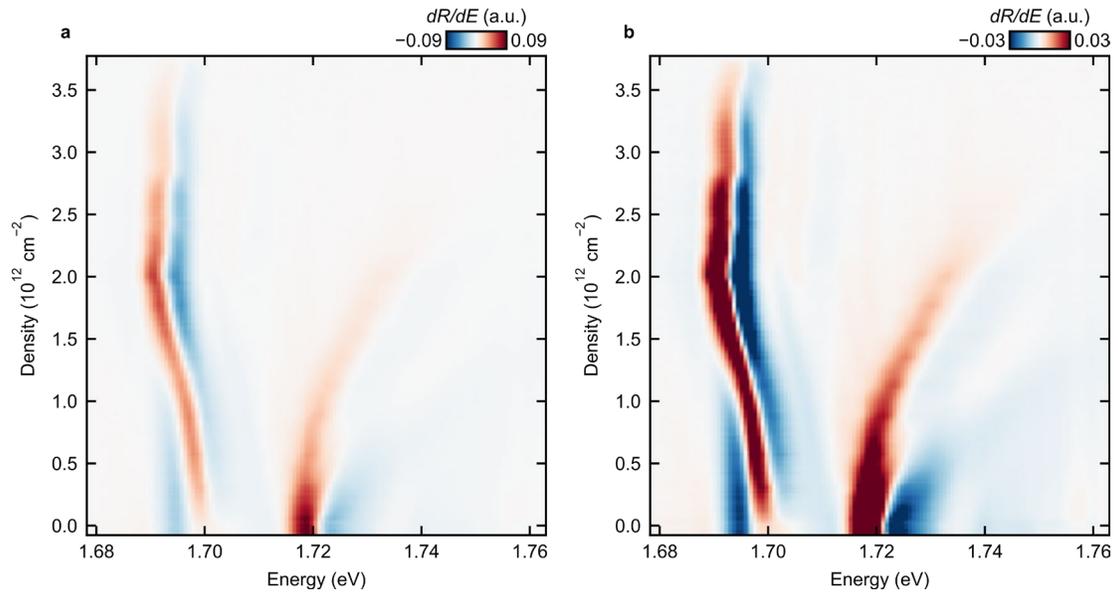

**Extended Data Fig. 2. | Pump-off gate-dependent dR/dE spectra at 9 T.** Panel b depicts the same data as in panel a, but with the color scale saturated.

**Extended Data Figure 3**

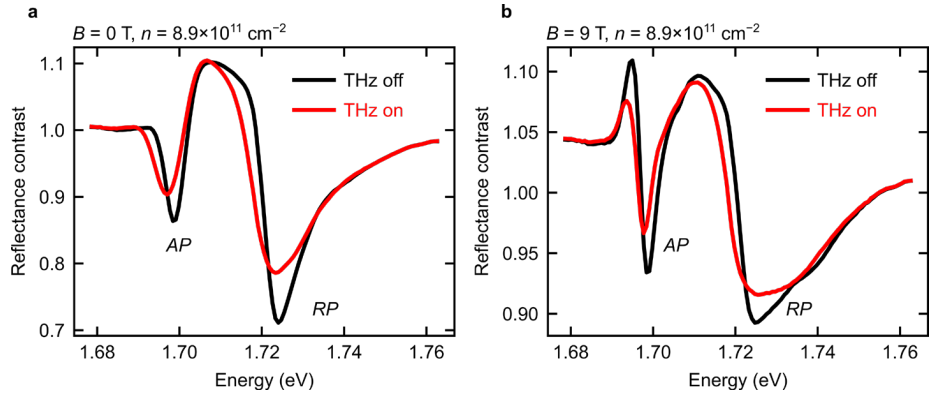

**Extended Data Fig. 3. | Raw reflectance spectra with THz field on and off. a,** reflectance spectra at zero magnetic field and at a density of $8.9\times10^{11}$ cm$^{-2}$ with THz field off (black) and on (red). Both spectra are normalized to the raw reflection measured at a high density of $3.4\times10^{12}$ cm$^{-2}$. **b**, reflectance spectra under a magnetic field of 9 T and at a density of $8.9\times10^{11}$ cm$^{-2}$ with THz field off (black) and on (red). Both spectra are normalized to the raw reflection measured at $3.4\times10^{12}$ cm$^{-2}$ and at fixed delays times of 0.8 ps (a) and −0.2 ps (b).

**Extended Data Figure 4**

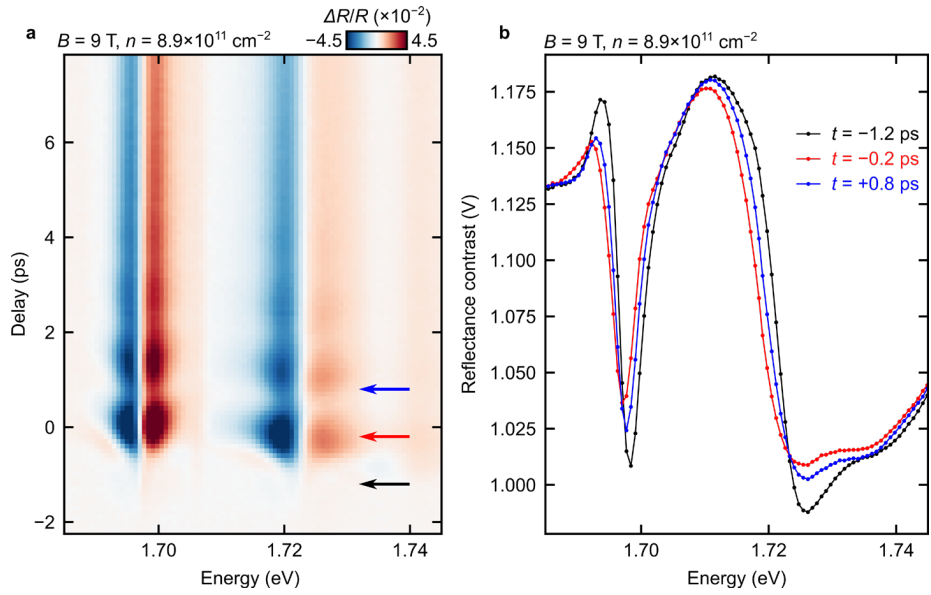

**Extended Data Fig. 4. | Time-dependent reflectance contrast spectra. a,** $\Delta R/R$ measured as a function of time at B=9 T and at a hole density of $8.9 \times 10^{11}$ cm$^{-2}$. The arrows indicate the delay times at which spectra are extracted in panel b. **b,** corresponding pump-on reflectance spectra at the delay times indicated in panel a.

**Extended Data Figure 5**

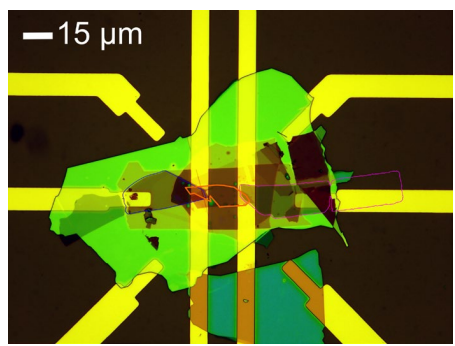

**Extended Data Fig. 5. | Device image.** The dual-gated region where WSe$_2$ and MoSe$_2$ overlap is highlighted in orange. The graphite contact to WSe$_2$ is highlighted in purple. The graphite contact to MoSe$_2$ is highlighted in blue.

**Extended Data Figure 6**

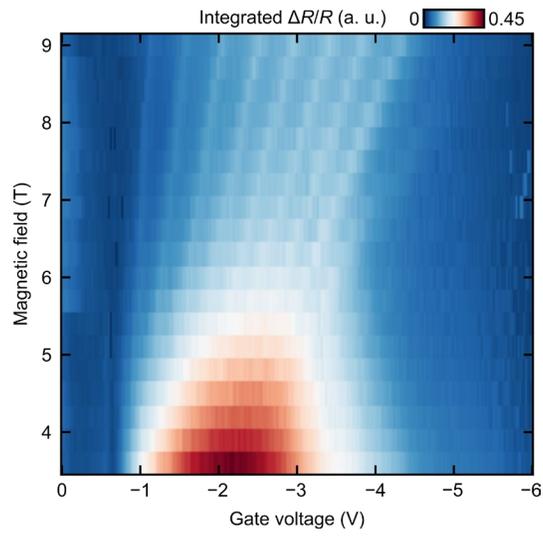

**Extended Data Fig. 6. | Spectrally integrated ΔR/R Landau fan for capacitance calibration.** At a delay of $\Delta t \approx 0.8$ ps, ΔR/R spectra are measured, and the integral of the absolute value of ΔR/R over the repulsive polaron resonance is calculated at each gate voltage (see Methods).

**Extended Data Figure 7**

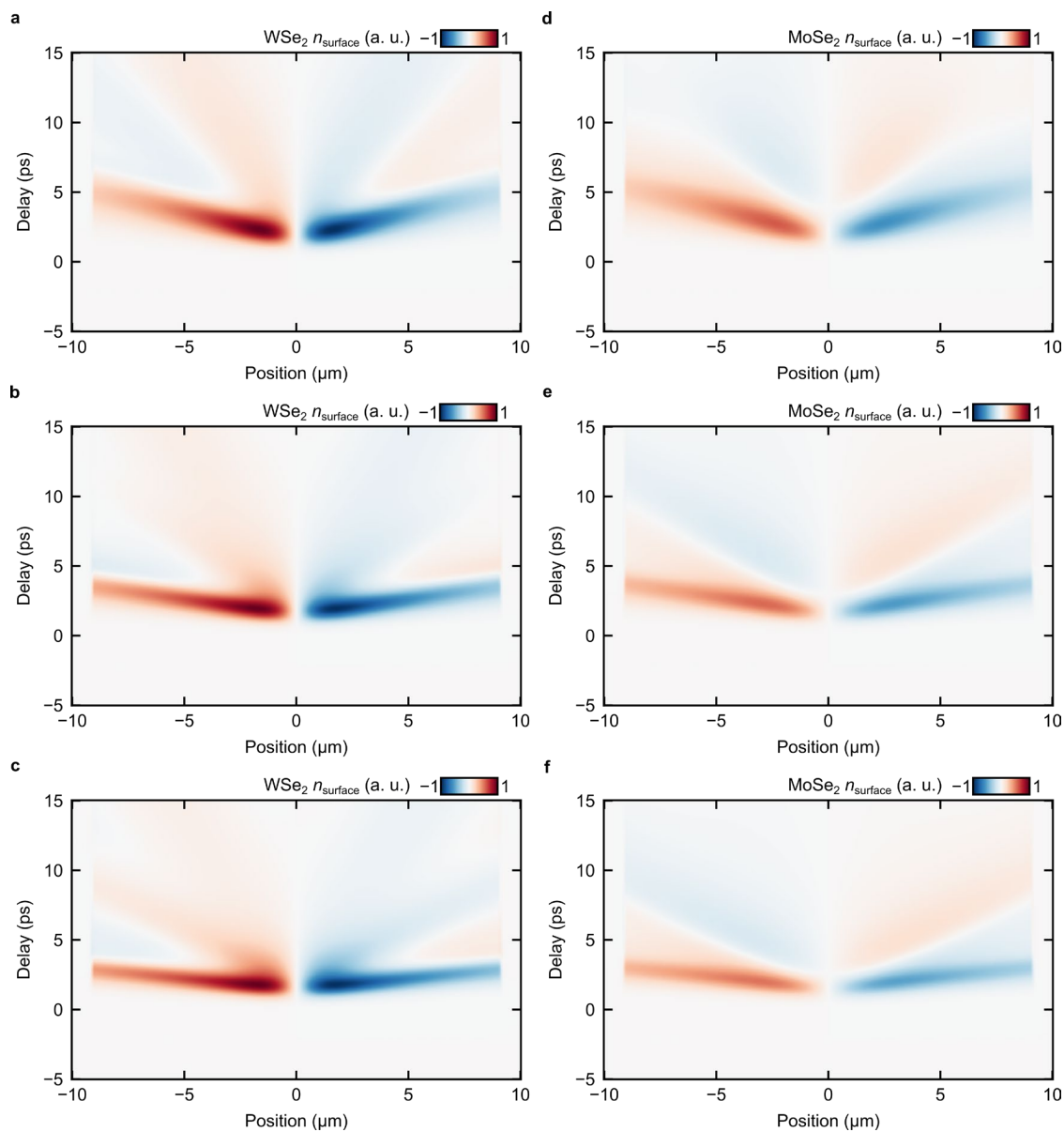

**Extended Data Fig. 7. | Simulated transients. a-c,** induced transient surface charge on WSe$_2$ at equilibrium densities of 2.9×10$^{11}$ cm$^{-2}$ (a), 7.7×10$^{11}$ cm$^{-2}$ (b) and 1.4×10$^{12}$ cm$^{-2}$ (c) simulated with FDTD (see Methods). The color scale on each is normalized to the overall maximum value of |$n_{\text{surface}}$(t)|. **d-f,** corresponding induced transient surface charge on MoSe$_2$ at equilibrium densities of 2.9×10$^{11}$ cm$^{-2}$ (d), 7.7×10$^{11}$ cm$^{-2}$ (e) and 1.4×10$^{12}$ cm$^{-2}$ (f) computed simultaneously with the panels on the left. The color scales on panels d-f are the same as those at the same density among panels a-c, reflecting the fact that the induced charge on MoSe$_2$ is smaller than that on WSe$_2$ (owing to the increased mass).